\documentclass{raa}

\usepackage{graphicx,times}             
\input{epsf.sty}                        

\begin{document}

   \title{The LEGUE High Latitude Bright Survey Design for the LAMOST Pilot Survey
}

   \volnopage{Vol.0 (200x) No.0, 000--000}      
   \setcounter{page}{1}          

   \author{Yueyang Zhang \inst{1,3}
   \and Jeffrey L. Carlin \inst{2}
   \and Fan Yang \inst{1,3}
   \and Chao Liu \inst{1}
   \and Licai Deng \inst{1}
   \and Heidi Jo Newberg \inst{2}
   \and Haotong Zhang \inst{1}
   \and S\'ebastien L\'epine \inst{4}
   \and Yan Xu \inst{1}
   \and Shuang Gao \inst{1}
   \and Norbert Christlieb \inst{5}
   \and Zhanwen Han \inst{6}
   \and Jinliang Hou \inst{7}
   \and Hsutai Lee \inst{8}
   \and Xiaowei Liu \inst{9}
   \and Kaike Pan \inst{10}
   and Hongchi Wang \inst{11}
   }

   \institute{Key Lab for Optical Astronomy, National Astronomical Observatories, Chinese Academy of Sciences,
             Beijing 100012, China ({\it zhangyy@bao.ac.cn})\\
        \and        Department of Physics, Applied Physics, and Astronomy, Rensselaer Polytechnic Institute, 110 8th Street, Troy, NY 12180, USA ({\it carlij@rpi.edu})\\
        \and        Graduate University of Chinese Academy of Sciences, Beijing 100049, China\\
        \and        Department of Astrophysics, Division of Physical Sciences, American Museum of Natural History, Central Park West at 79th Street, New York, NY 10024\\
        \and        University of Heidelberg, Landessternwarte, K\"onigstuhl 12, D-69117 Heidelberg, Germany\\
        \and        Yunnan Astronomical Observatory, Chinese Academy of Sciences, Kunming 650011, China\\
        \and        Shanghai Astronomical Observatory, Chinese Academy of Sciences, 80 Nandan Road, Shanghai 200030, China\\
        \and        Academia Sinica Institute of Astronomy and Astrophysics, Taipei, China\\
        \and        Department of Astronomy \& Kavli Institute of Astronomy and Astrophysics, Peking University, Beijing 100871, China\\
        \and        Apache Point Observatory, PO Box 59, Sunspot, NM 88349, USA\\
        \and        Purple Mountain Observatory, Chinese Academy of Sciences, Nanjing, Jiangsu 210008, China\\
   }

   \date{Received~~2012 month day; accepted~~2012~~month day}

   \abstract{ We describe the footprint and input catalog for bright nights in the LAMOST Pilot Survey, which began in October 2011.
     Targets are selected from two stripes in the north and south Galactic Cap regions, centered at
     $\delta=29^\circ$, with 10$^\circ$ width in declination, covering right
     ascension of 135$^\circ$-290$^\circ$ and -30$^\circ$ to 30$^\circ$ respectively. We selected spectroscopic targets from a combination of the SDSS and 2MASS point source catalogs. The catalog of stars defining the field centers (as required by the Shack-Hartmann wavefront sensor at the center of the LAMOST field) consists of all $V<8^m$ stars from the Hipparcos
     catalog. We employ a statistical selection algorithm that assigns priorities
     to targets based on their positions in multidimensional color/magnitude space. This scheme overemphasizes
     rare objects and de-emphasizes more populated regions of
     magnitude and color phase space, while ensuring a smooth, well-understood selection function. A demonstration of plate design
     is presented based on the Shack-Hartmann star catalog and an input catalog that was generated by our target selection routines. }

   \authorrunning{Y. Y. Zhang et al. }            
   \titlerunning{The Bright Survey Design of the LAMOST Pilot Survey }  
   \maketitle

%
%
\section{Introduction}           
\label{sect:intro}

Following a two-year commissioning period, a Pilot Survey with the
LAMOST telescope (also known as Guo Shou Jing telescope, GSJT; see Zhao et al. 2012 for an overview) began in October 2011. The LAMOST Pilot Survey, which will continue through the end of spring 2012, is an opportunity to test systems in survey mode while also obtaining valuable science data. The Pilot Survey is conducted in
preparation for the main LAMOST survey, which will begin in late 2012.
The survey consists of two main components: LEGUE (LAMOST Experiment
for Galactic Understanding and Exploration) will obtain spectra of millions of stars for the study of structure and substructure in the
Milky Way, and LEGAS (LAMOST ExtraGAlactic Surveys) is a
survey of galaxies and QSOs. The unique design of the telescope (Cui et al. 2012), with a $3.6-4.9$ meter aperture (depending on the direction in which the telescope is pointing) and a
$5^\circ$-diameter focal plane populated with 4000 robotically
positioned optical fibers, opens up new opportunities for large-scale
spectroscopic surveys (Zhao et al. 2012). This combination of a large field of view and
large aperture enables LAMOST to efficiently survey huge contiguous
areas of sky to faint magnitudes.

In this work, we focus on the Galactic structure portion of the
survey, or LEGUE. The LEGUE survey itself is split into two survey
modes -- one for observations on bright nights, and another on dark
nights -- with separate input target catalogs. The bright targets are also observed on dark/gray nights when the sky transparency is low. For the Pilot Survey,
bright nights are defined as $\pm5$ nights around the full moon, dark
nights as $\pm5$ nights around new moon, and grey time is in between.
The design of the dark nights survey, which focuses on faint stars to
study the Galactic halo, is discussed in Yang et al. (2012).
During the Pilot Survey, an additional six nights are set aside for
system engineering tasks, from 5-7 and 20-22 nights after new moon. Here we
discuss the survey that was designed to take advantage of the bright
and grey time during the LAMOST Pilot Survey.

On the bright nights,
relatively bright ($r \leq 16.5^m$) stars are observed in three diffentent regions of sky: a low-latitude
region near the Galactic anticenter (GAC), a Galactic disk region, and
a constant-declination stripe at $\delta \sim 29^\circ$.
In this paper we describe the design of the $\delta \sim
29^\circ$ stripe, leaving discussion of the disk survey design to
another work (Chen et al. 2012). Targets for the anticenter portion of the survey are being selected from a separate input catalog using data from the Xuyi photometric survey (Liu et al. 2012, in prep.), and will thus be described elsewhere. We discuss the selection of the
areas to be observed and the selection of targets based on both Sloan
Digital Sky Survey (SDSS; York et al. 2000) and Two Micron
All-Sky Survey (2MASS; Skrutskie et al. 2006)
photometry.

Bright targets
are observed when the moon is bright or when the atmospheric transparency is
poor. Due to poor weather conditions at the site, $\sim80\%$ of the telescope time is devoted to
observing bright targets (see Deng et al. 2012, Yao et al.
2012 for discussion of the site conditions and sample survey
strategies). During the Pilot Survey over 1~million spectra of bright
stars will be obtained, with more than 5~million bright targets to be
observed during the main LAMOST survey. It is thus important to test
targeting strategies for optimizing the selection of objects of
particular scientific interest to the collaboration, and to explore
the effectiveness of the Survey Strategy System (SSS) of LAMOST at
covering the sky uniformly. The Pilot Survey allows us to explore
these (and other issues) while simultaneously gathering data for
studies of Milky Way structure.

The structure of this paper is as
follows: Section 2 discusses the construction of the overall sky
catalog used for targeting and the selection of bright central guide
stars for active optics corrections, especially focusing on particular
elements that are unique to the bright nights survey.  The design and
target selection to generate the catalog to be input to the targeting
software is discussed in Section 3. Section 4 compares the input
magnitude and color distributions to those of the actual targets that
are fed to fibers in an observing plate. Finally, we conclude with a
brief summary of the bright nights survey design.




\begin{figure}[!t]
\centering
\includegraphics[
%
width=14cm,angle=0]{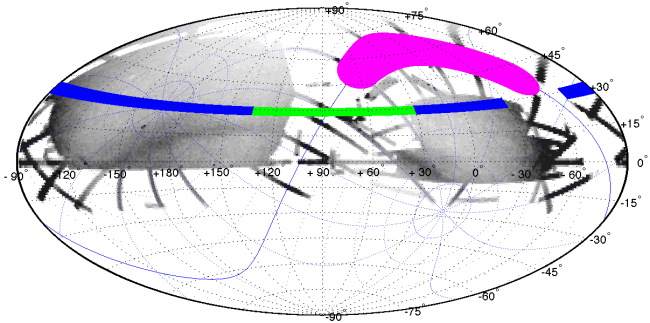}
\caption{The footprint of the LAMOST bright nights survey (colored regions) in equatorial coordinates overlaid on a starcount map from SDSS photometry. The magenta part is the disk survey region centered on the Galactic plane. The blue stripes are located in the North Galactic Cap (north stripe) and the South Galactic Cap (south stripe). The green stripe is the anticenter region at low Galactic latitudes ($\left| b \right| < 30^{\circ}$).}
\end{figure}

\section{Data and target selection}
\label{sect:data}

The footprint of the bright nights survey is shown in equatorial ($\alpha,
\delta$) coordinates in Figure~1. The magenta shape is the disk survey
region centered on the Galactic plane (i.e., $b=0^\circ$), which will
be discussed in another paper (Chen et al. 2012). The
blue and green stripes in Figure~1 are centered at a
declination of $29^\circ$. The green stripe is located at low ($b <
30^\circ$) Galactic latitudes in the Galactic anticenter region. This stripe is
mostly outside the SDSS footprint, and thus requires a different
source of photometry and astrometry for LAMOST target assignment.
Target selection for the anticenter region is being done based on
photometry from the Xuyi Schmidt Telescope Photometric Survey of the
Galactic Anti-center (XSTPS-GAC). The Xuyi survey provides uniform
photometry to $i\sim19^m$ in a $\sim3600$ deg$^{-2}$ region in the Galactic anticenter, from $150^\circ < l < 210^\circ$ and $-30^\circ < b < 30^\circ$.
LAMOST plates were designed differently in this anticenter region than
in the rest of the LEGUE survey; the target selection for the
anticenter portion of the LEGUE survey will be discussed elsewhere.
The blue stripes in Figure~1 are located in the northern Galactic Cap
(north stripe) and in the southern Galactic Cap (south
stripe). 
In this paper we discuss target selection for these two blue regions only; the disk and anticenter regions will be discussed elsewhere.

The constant declination stripe was placed at $\delta=29^\circ$ in part because the unique design of LAMOST
makes this the optimal direction to point the telescope in terms of
image quality. Mirror A, which acts as a Schmidt corrector
despite being the first element in the optical path, is the steerable
element that is pointed to targeted regions on the sky.
Mirror B (the spherical Schmidt ``primary'') is fixed at $25^\circ$
above the horizon.  Because the latitude at the site is $\sim40^\circ$, for Mirrors A and B to be aligned, Mirror A would have to be pointed to $\delta \sim -25^\circ$. Therefore, when Mirror A is pointed away from
$\delta \sim -25^\circ$ the effective collecting area decreases and the images become distorted. This means that
less light goes down the $3.3$-arcsecond fibers at higher declinations. In practice, the  telescope cannot point below $\delta = -10^\circ$. While this declination produces the best image quality based on the optics, it is at low altitude from the LAMOST site, making the atmospheric distortion significant. In practice, the optimal telescope performance is achieved at declinations near $25^\circ$.
The stripe at $\delta \sim 29^\circ$ also creates a region of contiguous observations that passes through
the Galactic anticenter when integrated with the (XuYi-selected)
anticenter catalogs and the LEGUE dark nights data. The width of the north
and south stripes is $10^\circ$, spanning $24^\circ < \delta <
34^\circ$. The right ascension range of the north stripe is $135^\circ -
290^\circ$, and the south stripe covers right ascensions from
$-30^\circ$ to $30^\circ$.

\begin{figure}[!t]
\centering
\includegraphics[width=12cm,angle=0]{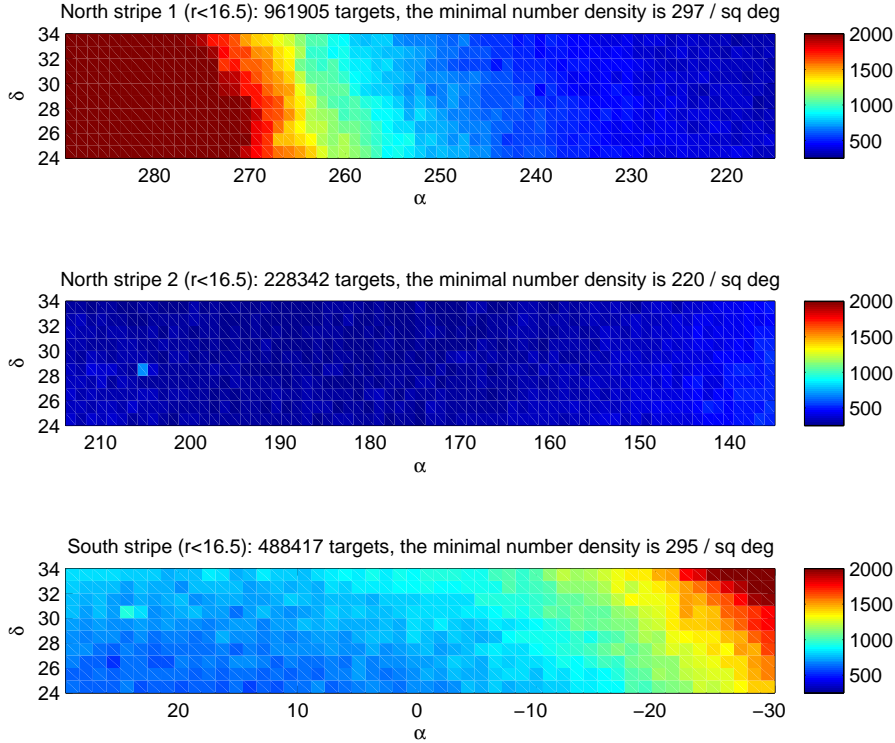}
\caption{The stellar number density of the north and south stripes, including all stars from 2MASS that do not have SDSS counterparts, as well as SDSS stars with $r < 16.5^m$. The number density in the north stripe is $\sim300$ deg$^{-2}$ in most of the sky near the North Galactic Pole. The lowest number density in the north stripe is 220 deg$^{-2}$. The number density in the south stripe is $\sim$600 deg$^{-2}$ in high latitude areas and can reach above 1500 deg$^{-2}$ near the plane. The lowest number density in the south stripe is 295 deg$^{-2}$.}
\label{fig:stripedensity_r_lt_16.5}
\end{figure}

Photometry from SDSS DR8 (Aihara et al. 2011) was used for
target selection in the blue stripes because it provides a uniform dataset covering the
entire region of interest. This choice was also motivated by a desire
to keep the bright and faint surveys as similar as possible; targets
for the LEGUE faint star survey (Yang et al. 2012)
were also selected from SDSS, so that when the bright and faint
surveys are combined, they will provide a relatively complete and
uniform survey. However, the bright magnitude limit at which SDSS
photometry saturates is $g \sim 14^m$ (Yanny et al. 2009). To
extend the survey to brighter magnitudes, providing more targets, we
supplemented the SDSS catalog with near-infrared photometry from the
2MASS point source catalog.


\begin{figure}[!t]
\centering
\includegraphics[width=12cm,angle=0]{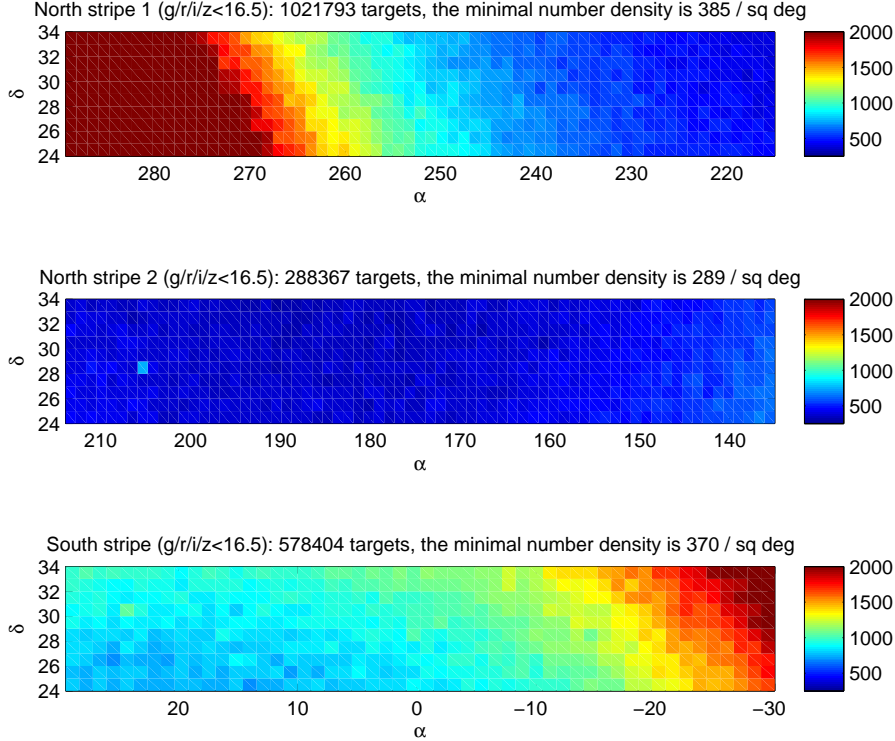}
\caption{Stellar number density in the north and south stripes including all stars with $g, r, i$, or $z$ magnitudes less than 16.5$^m$, and all 2MASS stars at the bright end. The number density in the north stripe (shown in the upper and middle panels) is $\sim400$ deg$^{-2}$ in most of the sky near the North Galactic Pole, with a minimum number density of 289 deg$^{-2}$. The number density in the south stripe (the bottom panel) is $\sim600$ deg$^{-2}$ in high latitude areas and can reach above 1500 deg$^{-2}$ at lower latitudes. The lowest number density in the south region is 370 deg$^{-2}$.}
\label{fig:stripedensity_all}
\end{figure}

The input catalog for the bright survey consists three subgroups
with different magnitude information: 1. targets with only SDSS
photometry; 2. targets with only 2MASS photometry; 3. targets with
both SDSS and 2MASS photometry. Initially, we intended to impose a
magnitude limit of $r < 16.5^m$ for targets on bright plates.
Figure~\ref{fig:stripedensity_r_lt_16.5} shows the density of targets
selected in this way for both the north and south stripes.
The number density of the north stripe is shown in the upper and middle panels of Figure~\ref{fig:stripedensity_r_lt_16.5} and the bottom panel shows the number density of the south stripe.
The color-coding in the figure represents the stellar density of available targets in each square degree of sky.
The total
stellar density including all 2MASS stars and $r < 16.5^m$ SDSS
candidates is less than 300 deg$^{-2}$ for much of the high-latitude
regions in the north stripe. The LAMOST fiber assignment program
typically requires three times the fiber density of candidates as
input in order to fill all of the available fibers; this requires an
input catalog of density 600 stars deg$^{-2}$. Obviously this density
is not achievable for bright stars at high latitudes.

LAMOST spectra cover a wide wavelength range of $3700 < \lambda <
9100$~\AA, so that some stars with extremely blue or red colors
that are fainter than magnitude 16.5 in $r$ will have enough flux to
obtain useful measurements in the blue or red regions of the spectra.
Thus, to increase the number of available targets for the bright
plates, we included stars with magnitudes brighter than 16.5$^m$ in $g, r,
i,$ or $z$-bands. For example, an M-dwarf with a red $r-i$ color of
$r-i = 1.2^m$ and $i = 16.3^m$ would have $r = 17.5^m$, but may be bright
enough to provide ample flux for measurement at red
(i.e., $i$ and $z$-band) wavelengths.  Thus the faint magnitude limit
employed for target selection in the bright survey plates is

\begin{equation}
  g < 16.5~|~r < 16.5~|~i < 16.5~|~z < 16.5
\end{equation}

\noindent where the ``$|$'' denotes ``or'' (i.e., only one or more of
these criteria must be met for a star to be included).

The faint magnitude limit for targets with only 2MASS photometry is
the same as the magnitude limit of 2MASS point source catalog:
\begin{equation}
  J < 15.8~\&~H <15.1~\&~K_s < 14.3.
\end{equation}

These magnitude selection criteria were chosen to ensure that the star
number density is high enough that most fibers can be occupied with targets, no matter where the field of view is placed. The lowest
number density in the north stripe, which appears near the northern
Galactic pole, is $\sim290$ deg$^{-2}$, while the lowest number
density in the south stripe is $\sim370$ deg$^{-2}$
(Figure~\ref{fig:stripedensity_all}).

\begin{figure}
\centering
\includegraphics[width=10cm,angle=0]{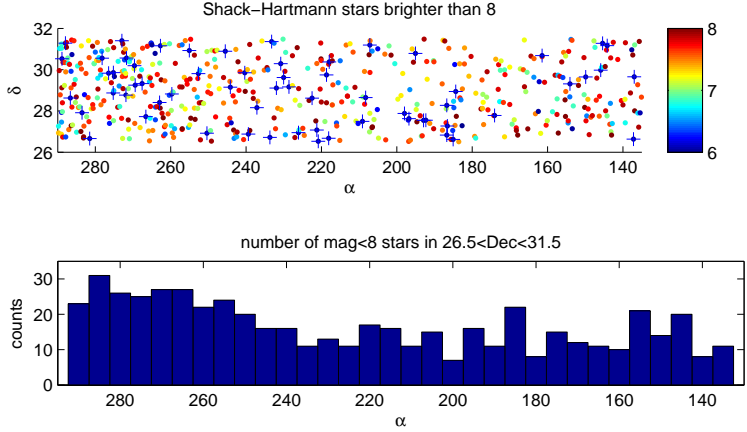}
\caption{The Shack-Hartmann star distribution of the north stripe along right ascension, considering stars with $V < 8^m$. Stars with $V<6^m$ are highlighted by ``+''. The lower panel shows that there are enough stars to make sure there are plates available to observe for a given right ascension. Note, however, that on bright nights when the sky background is elevated, the faint limit for Shack-Hartmann stars is between 6-7th magnitude, which limits the choice of available plate centers. }
\label{fig:SH_NGC}
\end{figure}

\begin{figure}
\centering
\includegraphics[width=10cm,angle=0]{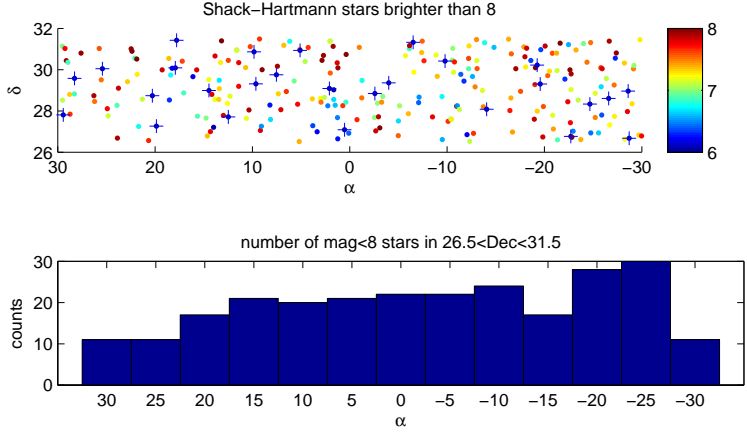}
\caption{As in Figure~\ref{fig:SH_NGC}, but showing the Shack-Hartmann star distribution of the south stripe along right ascension, considering stars with $V < 8^m$. There are more stars available in the southern sky, making the plate design more flexible in this region.}
\label{fig:SH_SGC}
\end{figure}

Active optics on LAMOST requires that each plate be
centered on a bright, $V < 8$ star that is fed to the Shack-Hartmann
wavefront sensor. For this purpose, we generated a catalog of
potential Shack-Hartmann stars from the Hipparcos catalog (Perryman et al. 1997). The
declination range of the Shack-Hartmann catalog is 26.5-31.5$^\circ$,
which is centered at the same declination ($29^\circ$) and covers half
the width of the north and south stripes. A field centered on a star with higher or lower declination would not fall completely within our desired observing footprint. The distribution of these stars
is seen in Figures~\ref{fig:SH_NGC}~and~\ref{fig:SH_SGC}, color-coded
by their Hipparcos $V$ magnitudes.
Stars with $V<6^m$ are highlighted by ``+''. For the bright survey, where plates
are often being observed in periods of elevated sky background (either
due to moonlight or clouds/haze), it is often necessary to use a
brighter central star (say, $V\sim6^m$) in order for the Shack-Hartmann
system to measure sufficient flux above background for image
correction. As can be seen in Figures~\ref{fig:SH_NGC} and
\ref{fig:SH_SGC}, limiting the central star to $V<6^m$ places strict
limits on the available plate centers.


The stripes and their corresponding Shack-Hartmann catalogs were
designed to cover a $10^\circ$ declination range to ensure that there
are always plates available to observe near a given right ascension.
Because LAMOST is a fixed meridian telescope, it is essential that
multiple plates are available at each right ascension for flexibility in plate selections.

The selection function can be assessed by the simulation of Galactic models like Sharma et al.'s synthetic survey in the future.

\section{sky to input catalog}
\label{sect:sky}

The LEGUE Pilot Survey target selection algorithm is presented in
detail in Carlin et al. (2012). Here we give a brief overview and
some details that are specific to the bright-star portion of the Pilot
Survey, but the global target selection routine is similar to that for
the faint Galactic halo portion of the Pilot Survey. The goal is to
overemphasize relatively rare stars in sparsely-populated regions of
parameter space while sampling large numbers of more common types of
stars.

The LAMOST fiber assignment algorithm of SSS performs optimally with an input
catalog of roughly three times the desired target density. There are
200 fibers per square degree in the LAMOST focal plane, so the input
catalogs should ideally contain at least 600 candidates per square
degree. As discussed in Section~\ref{sect:data} and seen in
Figure~\ref{fig:stripedensity_all}, this target density cannot be
achieved at high latitudes in a sample that is limited to magnitudes
brighter than 16.5$^m$. Thus we choose to create an input catalog that contains all
stars between $14^m < r < 16.5^m$, and all stars from 2MASS that do not
have an SDSS counterpart (i.e., mostly stars brighter than $r = 14^m$).
The catalog that is given to the fiber assignment program is generated
using the algorithm outlined in Carlin et al. (2012). This
algorithm selects candidates based on a general probability function

\begin{equation}
P_{\rm j, D} = \frac{K_{\rm D}}{[\Psi_0(\lambda_{\rm i})]_{\rm j}^{\alpha}} f_{\rm i} (\lambda_{\rm i})
\end{equation}

\noindent where $\lambda_{\rm i}$ denotes any observable (i.e.,
photometry, astrometry, or any combination of observed quantities) and
$\Psi_0(\lambda_{\rm i})$ is the statistical distribution function of
the observable $\lambda_{\rm i}$. The $K_{\rm D}$ term is a
normalization constant to ensure that the probabilities sum to one.
The $f_{\rm i} (\lambda_{\rm i})$ term is an optional function that
can be used to overemphasize targets in particular regions of
parameter space; this function is not used (i.e., $f_{\rm i}
(\lambda_{\rm i}) = 1$) in the bright survey. The density of stars is
calculated in multidimensional observable space, and the candidates
are weighted by a power of this ``local density''. The exponent
$\alpha$ that determines the weighting is typically between 0 and 1;
when $\alpha=0$, the probability of each target to be selected is the
same (i.e,. a random selection) and the distribution of the selected
sample will be the same as that of the input sample. When $\alpha=1$,
the probability of a target to be selected is inversely proportional
to the local density in the observable space, producing a selected
sample that is evenly distributed across the observable space. As a
result, the rare objects are over-emphasized. For the LEGUE bright
survey targets, we selected an intermediate case of $\alpha=1/2$
(i.e., weighting by the inverse square root of the local density),
which emphasizes the selection of rare objects but keeps a large
number of stars from higher-density regions in the observables. For
stars having SDSS magnitudes, we used $r, g-r$, and $r-i$ to calculate
the local density for targeting, and for those with 2MASS photometry only,
density was defined in $J, J-H, J-K$ parameter space. Unlike the dark
night survey of faint targets (Yang et al. 2012), we did not add any linear weights (i.e., $f_i=1$) in
color or magnitude to the bright survey selection criteria.

\begin{figure}[!t]
\centering
\includegraphics[width=10cm,angle=0]{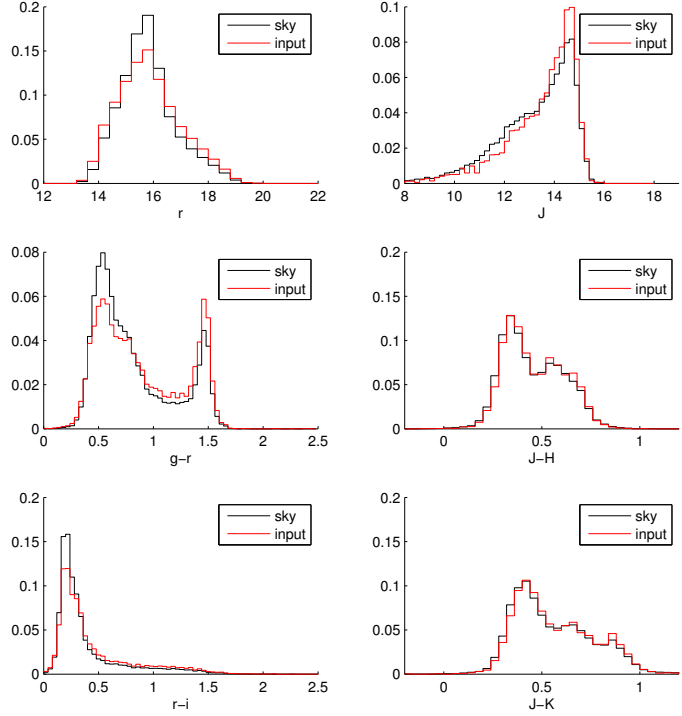}
\caption{Magnitude and color distributions of stars in a low Galactic latitude ($b \approx -20^\circ$) sky area and its corresponding input catalog. The left column depicts all stars with SDSS photometry, and the right includes stars with only 2MASS data. Black lines represent the distribution of all the original data in the sky area, and red lines are 600 candidates deg$^{-2}$ selected using the target selection method discussed in the text. Targets to be input to the fiber assignment program contain a smaller fraction of relatively common turnoff stars (at $0.3^m < g-r < 0.6^m$), with redder stars overemphasized. The 2MASS distributions of ``sky'' and ``input'' stars are similar because nearly all of the small number of available targets are selected by the targeting routine.}
\label{fig:skyInput low}
\end{figure}

Because the selection probabilities for stars appearing in SDSS were
generated separately from those appearing only in 2MASS, the two
resulting catalogs were concatenated and renormalized so that they sum
to one. The resulting catalog was then used to select target
candidates to give to the fiber assignment program.

As is evident in Figures~\ref{fig:stripedensity_r_lt_16.5}~and~\ref{fig:stripedensity_all}, the input catalog density at high Galactic latitudes is less than 400 deg$^{-2}$, even after including bright 2MASS targets with the SDSS stars. Thus, the selection of 600 stars per square degree is not possible at high latitudes, and all available stars will be included in the catalog input to the SSS fiber assignment program. At low latitudes, the high density of available targets means that we must sub-sample using the algorithm described above. Here we illustrate the effect of the target selection algorithm by comparing the magnitude and color distributions of the original data on the sky and the selected input catalog for a low-latitude ($\left| b \right| \sim 20^\circ$) field. This example field is at $330^\circ < \alpha < 336^\circ$ (i.e., $-30^\circ < \alpha < -24^\circ$), $28^\circ < \delta < 34^\circ$, with $b \approx -20^{\circ}$. Figure~\ref{fig:skyInput low} shows the
magnitude and color distributions of the original data and the
selected input catalog from the same sky area for this low-latitude field. Panels show stars with
SDSS photometry in the left column and those with only 2MASS photometry in the right column. The effect of the overemphasis on relatively rare stars is evident; this is especially clear in the middle left panel, which shows the $g-r$ color distribution. The selected candidates (the red histogram) have a lower peak near the main sequence turnoff locus ($g-r \sim 0.3-0.6^m$) than the overall distribution, a slight excess at intermediate colors ($0.6^m < g-r < 1.3^m$), and a clear excess of M-type stars at $g-r \sim 1.5^m$. The M-star excess arises because of the wide spread of M-dwarf $r-i$ colors at a nearly constant $g-r \sim 1.5^m$; this causes the density of M stars to be ``diluted'' in three-dimensional $r, g-r, r-i$ space (relative to the piling up in $g-r$ alone), giving the metal-poor tail (at red $r-i$ colors) of these relatively common stars a high probability of selection. There are only a small number of bright 2MASS candidates available outside of the Galactic disk region, so in most areas the 
``sky'' and ``input'' distributions in $J, J-H, J-K$ will look nearly identical.


\section{input catalog to plate design}
\label{sect:input}

\begin{figure}
\centering
\includegraphics[width=10cm,angle=0]{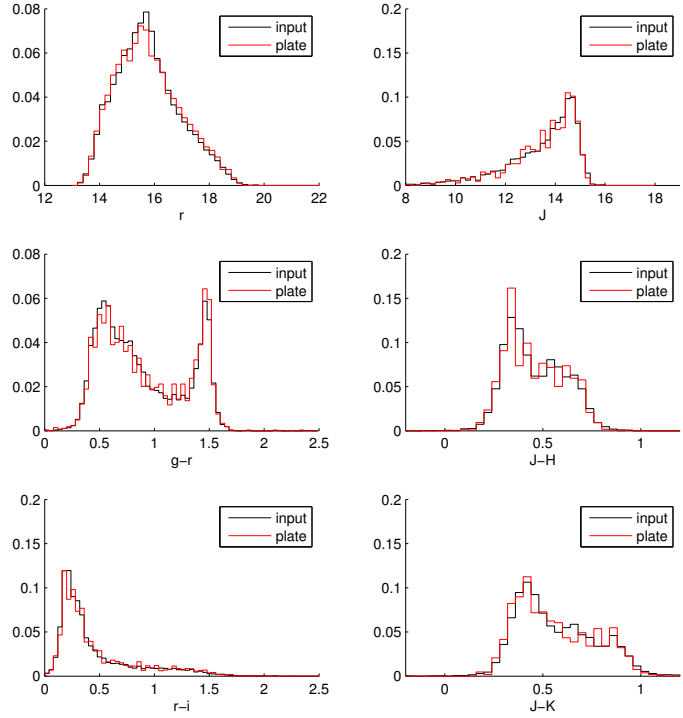}
\caption{ The magnitude and color distribution of the simulated plate used in Figure~\ref{fig:skyInput low} and its corresponding input catalog. The center of the simulated plate and the corresponding input catalog is ($\alpha, \delta$)
$ = (333^{\circ}, 31^{\circ}$). The plate is a circle with a radius of 2.5$^{\circ}$. The input catalog is a 6$^{\circ} \times 6^{\circ}$ rectangular box. Modulo some statistical fluctuations, the distribution of the targets selected for a plate are nearly identical to the input catalog. }
\label{fig:inputPlate_low}
\end{figure}

When SSS designs a plate for observations, it assigns targets of the
input catalog to fibers depending on priorities given to targets in the input catalog. For the Pilot Survey, priorities were assigned in a way that roughly reproduces the desired distribution of targets according to the assignment probabilities calculated for each star (for more details, see Carlin et al. 2012).
In principle, the magnitude and color distributions of a designed LAMOST plate
should be the same as those of the input catalog in the corresponding sky
area. To show this, we simulated the generation of a plate using SSS for the low-latitude field illustrated in Figure~\ref{fig:skyInput low}.
The center of the simulated plate
and the corresponding input catalog is $(\alpha, \delta) = (333^{\circ}$,
31$^{\circ}$). The plate is a circle with a radius of
2.5$^{\circ}$ centered on an available Shack-Hartmann star. The input catalog is a rectangular box with $330^\circ < \mathrm{RA}
< 336^\circ$, $28^\circ < \delta < 34^\circ$,
the same area as exemplified in Section 3.
Figure~\ref{fig:inputPlate_low} shows the input catalog as a solid black line, and the targets selected by SSS for the plate as a red line. Other than minor statistical fluctuations, this figure shows that the magnitude and color distribution of the
generated plate represents the input catalog well. We are thus confident that our target selection process is yielding the desired distribution of targets in the final survey design.

\begin{figure}
\centering
\includegraphics[width=\textwidth,angle=0]{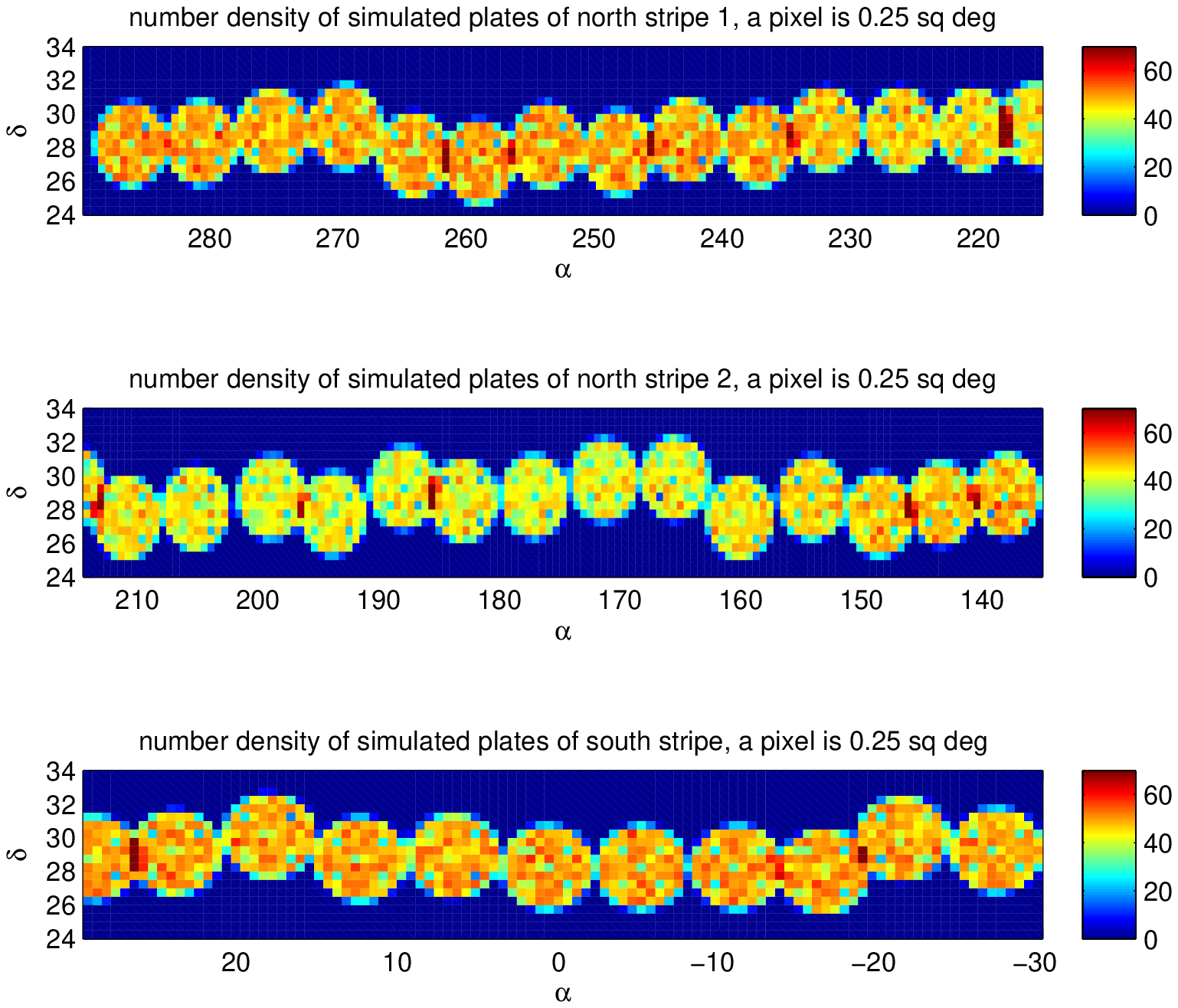}
\caption{A simple demonstration of how plates could be placed along the north and south stripes. Plates are generated by running a large input catalog based on our target selection criteria through a SSS simulation, with the center of each plate determined by the available Shack-Hartmann stars. The circles represent LAMOST plates, with the 0.25-degree squares giving the density of selected stars in each plate. Note that the density of stars in the NGC (upper panel) region drops with increasing latitude (roughly along $\alpha$ in this stripe), so that fewer than 200 stars deg$^{-2}$ are observed at high latitudes. }
\label{fig:demon plates}
\end{figure}

Based on the input catalog and the list of available Shack-Hartmann
stars, we use SSS to simulate a series of plates that covers the north and
south strips along $\delta \sim 29^\circ$. Plates are placed with a little overlap on the edge; in this simulation, meant for illustration only, there was little concern about optimally tiling regions of sky. Real
observational constraints and weather conditions are not considered. On a nightly basis, these will be
important factors in the design of plates due to the limited range about the meridian to which LAMOST can point. Figure~\ref{fig:demon plates} shows the results of this simple simulation -- the upper panel illustrates a string of plates in the north Galactic cap, and the lower panel, the southern cap. The figure shows the density of stars selected for observation by SSS in 0.25-degree squares on the sky. Because of the low number density of available stars in the north Galactic pole,
the mean density of those plates with high Galactic latitudes is $\sim$160
stars deg$^{-2}$, falling short of the 200 deg$^{-2}$ fiber density of LAMOST. The declining stellar density in the north strip is evident as a decreasing target density with increasing right ascension in the upper panel. Table 1 and Table 2 present the numbers of targets along with the coordinates of the plates' centers for the north and south stripe respectively.

\begin{table}
\begin{center}
\caption[]{The Numbers of Stars Targeted in the Plates Simulated along the North Stripe}\label{Tab:north_sim}

\begin{tabular}{cccc}
\hline\noalign{\smallskip}
plate No. & $ \alpha$ of the plate center & $\delta$ of the plate center & number of targets \\
\hline\noalign{\smallskip}
north 01 & 137.9892450 & 28.8715765 & 3685 \\
north 02 & 143.3263274 & 28.3680181 & 3637 \\
north 03 & 148.4965638 & 27.6954653 & 3566 \\
north 04 & 154.1171840 & 28.6827033 & 3494 \\
north 05 & 160.0882441 & 27.5252721 & 3416 \\
north 06 & 165.4555274 & 29.8712896 & 3321 \\
north 07 & 171.3191614 & 29.7587449 & 3293 \\
north 08 & 177.1524763 & 28.7999489 & 3259 \\
north 09 & 183.0040488 & 28.5364632 & 3253 \\
north 10 & 187.9606690 & 29.3141653 & 3256 \\
north 11 & 193.7844659 & 27.7665362 & 3244 \\
north 12 & 198.9016961 & 28.7419682 & 3270 \\
north 13 & 205.1630134 & 28.0651406 & 3348 \\
north 14 & 210.8639400 & 27.5099545 & 3443 \\
north 15 & 215.5584468 & 29.3699840 & 3440 \\
north 16 & 220.3432808 & 29.0592524 & 3525 \\
north 17 & 226.0668756 & 29.0785150 & 3584 \\
north 18 & 231.9577163 & 29.1054916 & 3684 \\
north 19 & 237.1434012 & 28.1567769 & 3751 \\
north 20 & 242.8491569 & 28.4387012 & 3793 \\
north 21 & 248.1120058 & 27.7094406 & 3771 \\
north 22 & 253.7299374 & 28.1370510 & 3791 \\
north 23 & 258.9954871 & 27.1343421 & 3767 \\
north 24 & 264.0892623 & 27.5668072 & 3718 \\
north 25 & 269.4409611 & 29.2479253 & 3805 \\
north 26 & 275.1055312 & 28.9835392 & 3757 \\
north 27 & 280.7169777 & 28.3027952 & 3792 \\
north 28 & 286.2285437 & 28.1882454 & 3792 \\
\noalign{\smallskip}\hline
\end{tabular}
\end{center}
\end{table}

\begin{table}
\begin{center}
\caption[]{The Numbers of Stars Targeted in the Plates Simulated along the South Stripe}\label{Tab:south_sim}
\begin{tabular}{cccc}
\hline\noalign{\smallskip}
plate No. & $ \alpha$ of the plate center & $\delta$ of the plate center & number of targets \\
\hline\noalign{\smallskip}
south 01 & 332.8464044 & 29.2546550 & 3767 \\
south 02 & 338.6976207 & 29.9574026 & 3778 \\
south 03 & 343.5475406 & 28.0166028 & 3760 \\
south 04 & 348.9428409 & 28.2479046 & 3791 \\
south 05 & 354.8782129 & 28.2459536 & 3789 \\
south 06 &   0.7101522 & 28.2518911 & 3788 \\
south 07 &   6.5113298 & 28.9405581 & 3783 \\
south 08 &  12.2903523 & 28.7193068 & 3804 \\
south 09 &  17.9149447 & 30.0897296 & 3810 \\
south 10 &  23.7578593 & 29.1007859 & 3789 \\
south 11 &  28.7807900 & 28.7979522 & 2914 \\
\noalign{\smallskip}\hline
\end{tabular}
\end{center}

\end{table}

\section{Summary}
\label{sect:summary}
In this paper, we described the input catalog for observations on bright nights of the LAMOST Pilot Survey. The sky coverage of the survey consists of a contiguous stripe at roughly constant declination of $\sim29^\circ$.
We discussed details of the plate design, which included a combination of SDSS and 2MASS photometry.
The input catalog of the bright nights survey consists of SDSS
stars brighter than 16.5$^m$ and all 2MASS point sources brighter than
the limiting magnitude ($r \sim 14^m$) of SDSS. The target selection method is based on
pre-assigned priorities, which weight stars by the inverse square root of the ``local density'' in $r, g-r, r-i$ (or $J, J-H, J-K$ for those having only 2MASS magnitudes) to overemphasize rare objects and
de-emphasize the objects in more populated regions of magnitude and color phase
space. We illustrate the differences between the magnitude and color distributions of stars selected by our target selection method and the overall distribution for a given region of sky.

Overall, the LAMOST/LEGUE Pilot Survey will obtain 1-2 million stellar spectra of bright stars (in addition to the faint stars from the other components of the survey) spanning a range of Galactic latitudes. These spectra will yield numerous scientific results while also providing valuable test data for refinement of LAMOST survey operations.


\begin{acknowledgements}
This work is partially supported by National Natural Science Foundation of China (NSFC) through grant No.10573022, 10973015, 11061120454 and by Chinese Academy of Sciences (CAS) through grant GJHZ20081, and the US National Science Foundation through grant AST-09-37523.
\end{acknowledgements}

\label{lastpage}


\begin{thebibliography}{99}


\bibitem[Aihara et al. (2011)]{Aihara11} Aihara, H., Allende Prieto, C., An, D., et al.\ 2011, \apjs, 193, 29

\bibitem[Carlin et al. (2012)]{Carlin12} Carlin, J.~L., L\'epine, S., Newberg, H.~J., et al.\ 2012, \raa, in press

\bibitem[Chen et al. (2012)]{Chen12} Chen, L., Hou, J., Yu, J., et al.\ 2012, \raa, in press

\bibitem[Cui et al. (2012)]{Cui12}  Cui, X. Q., Zhao, Y. H., Chu, Y. Q., et al. 2012, \raa, in press

\bibitem[Deng et al. (2012)]{Deng12} Deng, L., Newberg, H.~J., Liu, C., et al.\ 2012, \raa, in press

\bibitem[Perryman et al. (1997)]{Perryman97} Perryman, M. A. C., Lindegren, L., Kovalevsky, J., et al.  1997, \aap, 323, 49

\bibitem[Sharma et al. (2011)]{Sharma11} Sharma, S.,
Bland-Hawthorn, J.,  Johnston, K. V., et al.\ 2011, \apj, 730, 3

\bibitem[Skrutskie et al. (2006)]{Skrutskie06} Skrutskie, M.~F.,
Cutri, R.~M., Stiening, R., et al.\ 2006, \aj, 131, 1163

\bibitem[Yang et al. (2012)]{Yang12} Yang, F., Carlin, J.~L., Liu, C., et al.\ 2012, \raa, in press

\bibitem[Yanny et al. (2009)]{yanny09} Yanny, B., Rockosi, C.,
Newberg, H.~J., et al.\ 2009, \aj, 137, 4377

\bibitem[Yao et al. (2012)]{Yao12} Yao, S., Liu, C., Zhang, H.~T., et al.\ 2012, \raa, in press

\bibitem[York et al. (2000)]{york00} York, D.~G., Adelman, J.,
Anderson, J.~E., Jr., et al.\ 2000, \aj, 120, 1579

\bibitem[Zhao et al. (2012)]{Zhao12} Zhao, G., Zhao, Y., Chu, Y., et al.\ 2012, \raa, in press


\end{thebibliography}
\end{document}